\documentclass[pdflatex,sn-apa]{sn-jnl}
\usepackage{graphicx}
\usepackage{amsmath,amssymb}
\usepackage{booktabs}
\usepackage{subcaption}

\begin{document}

\title[Textual Expansion and Team Size Changes]
{Changes in Manuscript Length, Research Team Size, and International Collaboration in the Post-2022 Period: Evidence from \textit{PLOS ONE}}

\author*[1]{\fnm{Yossi} \sur{Ben-Zion}}\email{benzioy@biu.ac.il}
\author[1]{\fnm{Eden} \sur{Cohen}}
\author[2]{\fnm{Nitza} \sur{Davidovitch}}

\affil*[1]{\orgdiv{Department of Physics},
\orgname{Bar-Ilan University},
\orgaddress{\city{Ramat Gan}, \postcode{52900}, \country{Israel}}}

\affil[2]{\orgdiv{Department of Education},
\orgname{Ariel University},
\orgaddress{\city{Ariel}, \postcode{40700}, \country{Israel}}}

\abstract{
Large language models (LLMs) have diffused rapidly into academic writing since late 2022. Using the complete population of 109,393 research articles published in \textit{PLOS ONE} between 2019 and 2025, we examine population-level structural publication indicators, including full-text manuscript length, authorship team size, reference volume, and cross-linguistic collaboration, before and after 2022. \textit{PLOS ONE}'s multidisciplinary scope and consistent editorial framework allow cross-field comparison under uniform conditions over an extended period. Manuscript length increased substantially, with gains ranging from 14.8\% among African-affiliated authors and 11.7\% among Asian-affiliated authors to 5.3\% among native English-speaking (NES) authors, cutting the word-count gap by 39\%. More strikingly, non-native English-speaking (NNES) authors reduced both authorship team size, from 6.54 to 6.06 authors, or 7.3\%, and collaboration with NES co-authors, from 17.8\% to 12.2\%, or 36\%, while NES authors remained stable in both team size and collaboration rates. Reference counts increased modestly and uniformly across groups. These findings suggest that post-2022 tools may be reshaping not only how science is written, but who writes it together.
}

\keywords{Manuscript length, Research team size, Scientific collaboration, Authorship patterns, Bibliometrics}

\maketitle

\section{Introduction}

For decades, the dominance of English as the primary language of international scientific publications has structured participation in global knowledge production. Although a majority of publishing scientists are non-native English speakers (NNES), the overwhelming share of indexed scientific output is written in English \citep{liu2017changing,kourilova2012some}. NNES scholars invest substantially more time in writing and revision \citep{amano2023manifold,ramirez2020disadvantages}, and face higher rejection rates attributable to language-related factors \citep{lenharo2023true}.

The public release of large language models (LLMs), most notably ChatGPT in late 2022, coincided with the introduction of widely accessible tools for drafting and language refinement in academic writing. Experimental evidence shows that generative AI increases writing productivity and improves perceived output quality \citep{noy2023experimental}. Bibliometric analyses document rapid diffusion of LLM-assisted writing across scientific domains, with adoption particularly pronounced among NNES scholars \citep{liang2024mapping,liu2025ai,filimonovic2025generative}. Difference-in-differences analyses further indicate post-2022 increases in lexical complexity in NNES-authored abstracts \citep{lin2025chatgpt}.

Bibliometric research uses indicators such as manuscript length, authorship team size, reference volume, and collaboration patterns to characterize the structure of scientific production \citep{moed2004handbook}. Whether the post-2022 period is associated with shifts in these structural publication characteristics, and whether such shifts differ between NES and NNES scholars, remains unexamined. Shifts in these indicators may have implications for participation and stratification within the scientific system.

We analyze trends in manuscript length, authorship team size, reference volume, and international collaboration before and after 2022 using a large-scale bibliometric dataset. We do not attempt to detect AI-generated text or infer tool usage at the document level. Instead, we examine objective publication indicators derived from complete full-text records. Our analysis focuses on the full population of research articles published in the multidisciplinary journal \emph{PLOS ONE} between 2019 and 2025 and indexed through the fourth quarter of 2025 at the time of data retrieval, enabling cross-field comparison within a consistent editorial framework. We treat 2022 as the final pre-diffusion benchmark year and compare subsequent publication characteristics to assess temporal shifts and differences across linguistic groups.

Accordingly, we address three research questions:
\begin{itemize}
    \item RQ1: Did manuscript length change following 2022, and did trends differ by linguistic background?
\item RQ2: Did authorship team size and international collaboration patterns shift across the same period, particularly among NNES scholars?
\item RQ3: Did reference volume exhibit differential temporal trends across linguistic groups?
\end{itemize}

\section{Methods}

\subsection{Data and Sample}

\textit{PLOS ONE} was selected for three reasons: its multidisciplinary scope enables cross-field comparison under uniform editorial conditions; its open-access policy and public API provide complete full-text records across the entire study period; and its consistently high submission volume ensures adequate sample sizes across linguistic and geographic subgroups in every year examined.

The dataset comprises 109,393 research articles published in \emph{PLOS ONE} between 2019 and 2025 and indexed as of November 2025 at the time of data retrieval. Table~\ref{tab:counts} reports article counts by year. Only articles classified as \textit{research articles} were retained; corrections, retractions, and errata were excluded, as their word counts reflect modified or withdrawn content rather than complete manuscripts.

Articles were retrieved programmatically via the \emph{PLOS ONE} public API using a custom Python script. For each calendar year, an initial query returned the total number of qualifying articles. Subsequently, metadata including DOI, title, article type, and subject classifications were retrieved in batches of 1,000 records per request. Full-text XML files were subsequently downloaded in groups of four articles per request, a batch size chosen to balance retrieval speed and server load.

Article type was verified using both the API metadata and the XML file, with the XML serving as a fallback when API metadata were missing or inconsistent.
\begin{table}[ht]
\centering
\caption{Number of research articles published in \emph{PLOS ONE} by year (2019–2025).}
\label{tab:counts}
\begin{tabular}{lc}
\hline
Year & Number of articles (N) \\
\hline
2019 & 15,300 \\
2020 & 16,015 \\
2021 & 16,094 \\
2022 & 15,584 \\
2023 & 13,814 \\
2024 & 16,203 \\
2025 & 16,383 \\
\hline
Total & 109,393 \\
\hline
\end{tabular}
\end{table}
\subsection{Manuscript Length}
Word counts were derived from the full-text XML of each article, including the front matter, body, and back matter. Extracted text was cleaned of residual markup and concatenated into a single string prior to tokenisation.

Words were defined as whitespace-delimited tokens of two or more characters. This threshold excludes single-character formatting artefacts (e.g., table border characters) that would otherwise inflate counts. The omission of single-character words (e.g., “I”, “a”) has negligible impact on cross-group comparisons. Word counts include the reference list.

Alternative counting specifications that varied with respect to the inclusion of references and single-character tokens were evaluated. The selected approach yielded the smallest deviation from manual counts. The procedure was validated on a random sample of 100 articles by comparison with Microsoft Word and several independent online word-counting tools. Observed deviations were negligible.

\subsection{References}

Reference counts were derived from the full-text XML of each article by counting bibliographic entries listed in the reference section.
\subsection{Author Language Background Classification}
Author names and institutional affiliations were extracted from the XML records. Each author was linked to their affiliation via unique bracket-enclosed identifiers (e.g., \texttt{[aff001]}), enabling accurate assignment when authors held multiple affiliations. Editor affiliations appearing in a subset of XML files were identified and removed prior to classification.

Language background was classified using GPT-5 mini via the OpenAI API. For each article, the model received the list of authors linked to their institutional affiliations and was asked three binary questions: (1) whether the first author was affiliated with an institution in a native English-majority country; (2) whether the last author was; and (3) whether at least one author was. Native English-majority countries were defined as the United States, United Kingdom, Ireland, Australia, New Zealand, and Canada. When an author held multiple affiliations, the article was coded as 1 if any affiliation met the criterion. Articles were processed in batches of up to 150 per API call. The full prompt is provided in Appendix~A.

Throughout all analyses, articles were classified according to the 
linguistic background of the first author (NES or NNES), who is typically the primary contributor to the research \citep{tscharntke2007author,lariviere2016contributorship}. In 92.8\% of articles in our dataset, the first 
and last authors belonged to the same linguistic group, indicating 
that classification by last author would yield substantively 
identical results.

To assess classification accuracy, 100 randomly selected articles were independently coded by the authors and compared to the model output. Agreement was 100\%.
This high rate reflects two factors. First, the classification task is relatively unambiguous: institutional affiliation with English-majority countries can typically be determined directly from the affiliation text. Second, the prompt was developed through iterative refinement until a stable formulation was achieved, as described in Appendix A. Manual inspection was applied throughout this development process. Critically, the validation sample was drawn after the final prompt was fixed, ensuring that the reported agreement reflects out-of-sample performance rather than in-sample optimization.

To further assess the robustness of the LLM-based classifications, we reprocessed a random sample of 1,000 articles on December 1, 2025, using three GPT-5 model variants available via the OpenAI API at the time of analysis (\texttt{GPT-5}, \texttt{GPT-5-mini}, and \texttt{GPT-5-nano}). These models were selected as they represented the most advanced and up-to-date iterations of OpenAI’s language models at the time of the study. The classifications were identical across all three models (100\% agreement). Raw outputs were stored prior to post-processing to enable verification and auditing.
\subsection{Geographic Classification}

Geographic classification was performed using a parallel large language model prompt (Appendix A), applied to author names and institutional affiliations. For each article, the model identified (1) the continent of the first author, (2) the continent of the last author, and (3) the presence of at least one author from each of five regions: Asia, Europe, Africa, the Americas, and Other (including Australia and New Zealand). For consistency with the language background classification, 
geographic region was assigned based on the institutional 
affiliation of the first author.

When an author held multiple affiliations spanning different continents, the continent of the first-listed affiliation was used.

\subsection{Field Classification}

Subject classifications were retrieved from the \emph{PLOS ONE} API in hierarchical format, where each subject tag represents a path from a broad domain to a specific sub-topic (e.g., \texttt{/Medicine and health sciences/Infectious diseases/Bacterial infections}). For each subject tag, only the top-level domain was retained. Across the study period, eleven distinct top-level domains were observed.

To assign each article a single primary field, an annual weighting procedure was applied. For each year $y$, let $N_y$ denote the total number of articles and $N_{iy}$ the number of articles in which domain $i$ appeared at least once. Each domain received a year-specific weight inversely proportional to its prevalence:
\begin{equation}
    w_{iy} = \frac{N_y}{N_{iy}}.
\end{equation}
For a given article $d$ published in year $y$, a weighted score for each domain $i$ was computed as:
\begin{equation}
    s_{iy}(d) = f_{iy}(d) \cdot w_{iy},
\end{equation}
where $f_{iy}(d)$ denotes the frequency with which domain $i$ appears among the article's subject tags. Duplicate subject tags were retained, as their repetition reflects the degree to which a domain characterises the article. The primary field was then assigned as:
\begin{equation}
    \operatorname{Field}(d) = \underset{i}{\arg\max} \; s_{iy}(d).
\end{equation}
This procedure assigns greater weight to less prevalent domains while preserving within-article subject emphasis through tag frequency.

\section{Empirical Strategy}

Because manuscript word counts were right-skewed, all primary analyses of manuscript length were conducted on log$_{10}$-transformed word counts. Figures are presented on the original scale for interpretability.

\paragraph{Temporal differences in manuscript length (2019--2025).}
Differences across publication years were assessed using one-way ANOVA on log$_{10}$(word count). Pairwise post-hoc comparisons were computed using estimated marginal means with Tukey adjustment.

As a robustness check, we additionally conducted a Kruskal--Wallis test on the original word-count scale with Dunn post-hoc comparisons (Bonferroni adjustment). These non-parametric tests yielded substantively identical conclusions and are therefore not reported.

\paragraph{Heterogeneity by first-author language background (2019--2025).}
To assess whether temporal patterns differed by first-author language background, we estimated a two-way ANOVA on log$_{10}$(word count) including main effects for year and first-author group and their interaction. Post-hoc contrasts were computed using estimated marginal means.

\paragraph{Heterogeneity by continent (2022 vs.\ 2025).}
For continent-level analyses, we restricted the sample to 2022 and 2025 and estimated a two-way ANOVA on log$_{10}$(word count) including year, continent group, and their interaction. Holm-adjusted post-hoc contrasts were computed; inference is based primarily on the interaction effects.

\paragraph{Heterogeneity by primary field (2022 vs.\ 2025).}
Field heterogeneity was assessed on the 2022/2025 subsample using a two-way ANOVA on log$_{10}$(word count) including year, primary field, and their interaction. Holm-adjusted post-hoc contrasts were computed; inference is based primarily on the interaction effects.

\paragraph{Difference-in-differences regression (2022 vs.\ 2025).}
Differential change in manuscript length between 2022 and 2025 by first-author language background was estimated using linear regression models on the 2022/2025 subsample. We estimated (i) a baseline model including year only, (ii) a difference-in-differences specification including the year $\times$ first-author interaction, and (iii) a full specification adding primary-field fixed effects and year-by-field interactions.

The models were estimated both on the original word-count scale and on log$_{10}$(word count). Results were substantively equivalent across specifications. For consistency with distributional properties and the ANOVA analyses, only the log$_{10}$ results are reported.

\paragraph{Authorship team size and reference counts (2022 vs.\ 2025).}
Authorship team size and reference counts were modelled using negative binomial regression with a log link, including the interaction between year and first-author language background. Exponentiated coefficients are reported as incidence rate ratios.

\paragraph{Collaboration with NES co-authors (2019--2025).}
Among articles with NNES first authors, the probability of having at least one NES co-author was modelled using logistic regression with publication year indicators and 2019 as the reference year. Results are reported as odds ratios with confidence intervals.

\subsection{Software and missing data.}
All statistical analyses were conducted in R (version 4.5.0). Linear and ANOVA models were estimated using base R functions; negative binomial regressions were estimated using the \texttt{MASS} package; post-hoc contrasts were computed using the \texttt{emmeans} package. Figures were generated in \textit{Python} (using \texttt{pandas} and \texttt{matplotlib}). Observations with missing values in the outcome or key predictors were excluded from the relevant model (complete-case analysis).

\section{Results}

\subsection{Temporal Shifts in Manuscript Length (2019--2025)}

\begin{figure}[htbp]
    \centering
    \begin{subfigure}[b]{0.48\textwidth}
        \centering
        \includegraphics[width=\textwidth]{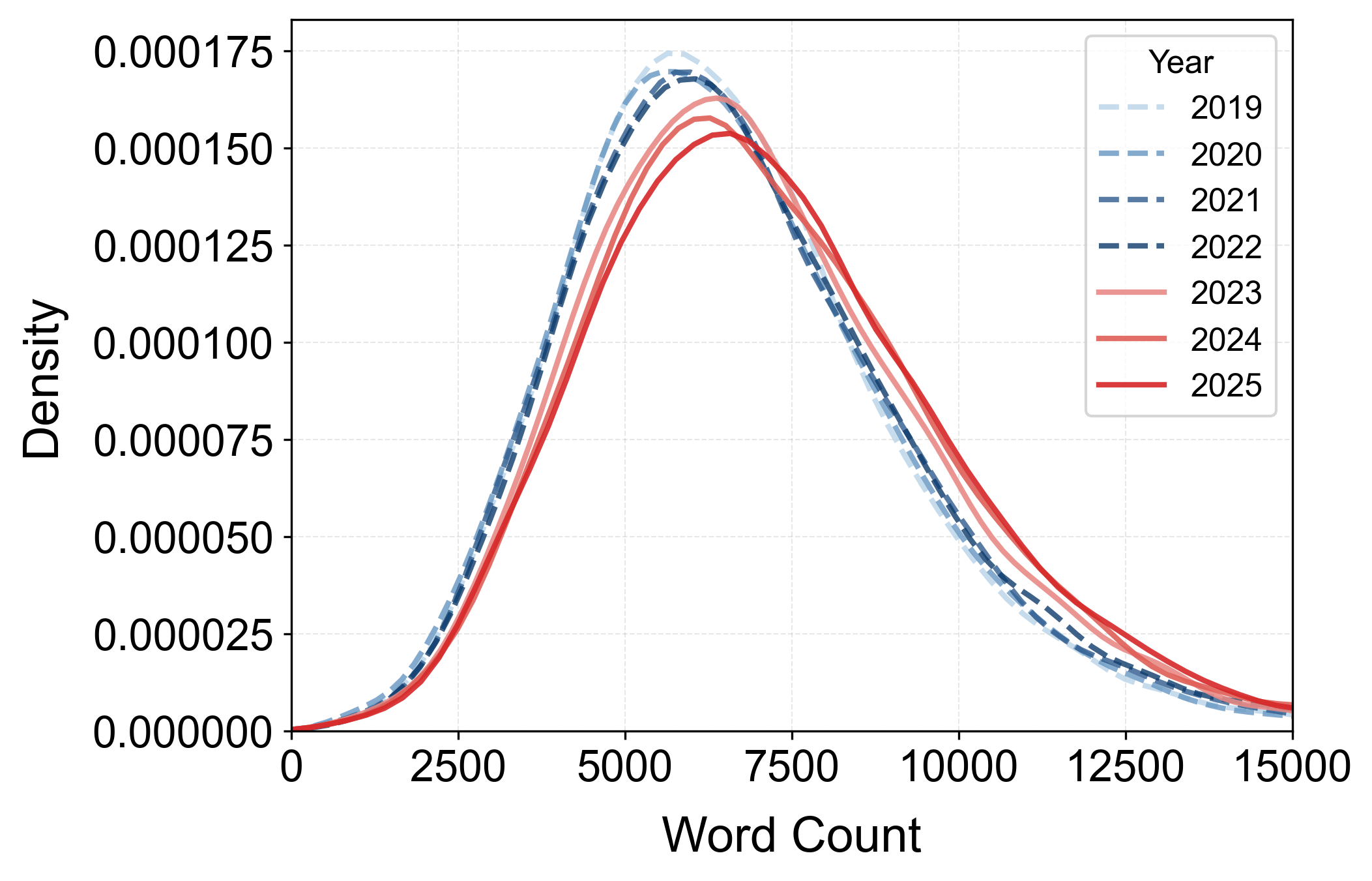}
        \caption{}
    \end{subfigure}
    \begin{subfigure}[b]{0.48\textwidth}
        \centering
        \includegraphics[width=\textwidth]{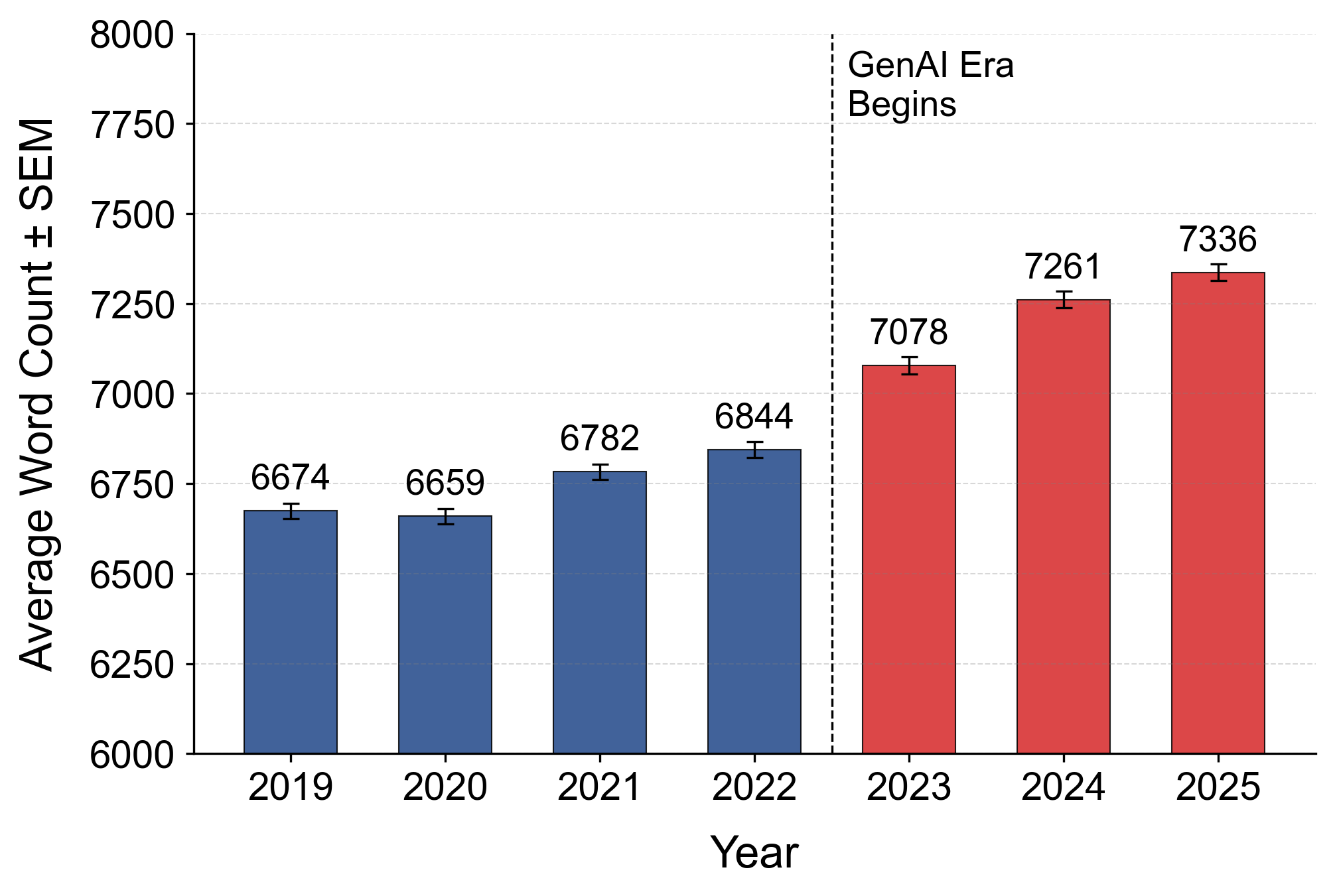}
        \caption{}
    \end{subfigure}
    \caption{\textbf{Manuscript length trends (2019--2025).}
    \textbf{(A)} Word count distributions by year (in words). Dashed lines represent 2019--2022; solid lines represent 2023--2025.
    \textbf{(B)} Mean article length (in words, $\pm$ SEM). The vertical dashed line indicates the onset of the significant increase observed in 2023.
    Statistical tests were conducted on log$_{10}$-transformed values; plotted values are shown on the original word-count scale.}
    \label{fig:length_trends}
\end{figure}

Article length increased non-linearly between 2019 and 2025. Because word counts were right-skewed, all statistical analyses were conducted on log$_{10}$-transformed values; figures display values on the original word-count scale for interpretability (Figure~\ref{fig:length_trends}).

Figure~\ref{fig:length_trends}A reveals two distinct periods in the distribution of article lengths: between 2019 and 2022, the distributions largely overlap, whereas from 2023 onward a rightward shift is evident, indicating a higher overall level of manuscript length compared to the pre-2023 period.

A one-way ANOVA confirmed a significant effect of publication year on log$_{10}$(word count) ($F(6, 109{,}386) = 145.80$, $p < .001$). Post-hoc comparisons (Tukey HSD) showed no difference between 2019 and 2020 ($p = .897$). Word counts increased between 2020 and 2021 ($p = .001$), though the cumulative difference from 2019 to 2021 did not reach statistical significance ($p = .082$). No change was observed between 2021 and 2022 ($p = .373$). A pronounced increase occurred between 2022 and 2023 ($p < .001$), representing the largest annual shift observed during the study period, followed by a further increase in 2024 relative to 2023 ($p < .001$). No significant difference was detected between 2024 and 2025 ($p = .318$).

Consistent with these patterns, mean article length increased sharply beginning in 2023 (Figure~\ref{fig:length_trends}B). By 2025, articles were on average approximately 662 words longer than in 2019, and all comparisons with the 2019 baseline from 2022 onward were statistically significant (all $p < .001$).

\subsection{Differential Acceleration and Linguistic Convergence}

Aggregated trends concealed substantial heterogeneity by authors' language background. We therefore stratified analyses by whether the first author was a NES or NNES.

A two-way ANOVA on log$_{10}$(word count), with year and first-author language background as factors, revealed significant main effects of year ($F(6, 109{,}375) = 147.25$, $p < .001$) and language background ($F(1, 109{,}375) = 1{,}129.45$, $p < .001$), as well as a significant interaction ($F(6, 109{,}375) = 11.01$, $p < .001$), indicating that temporal changes in manuscript length differed between groups.

\begin{figure}[htbp]
    \centering
    \begin{subfigure}[b]{0.48\textwidth}
        \centering
        \includegraphics[width=\textwidth]{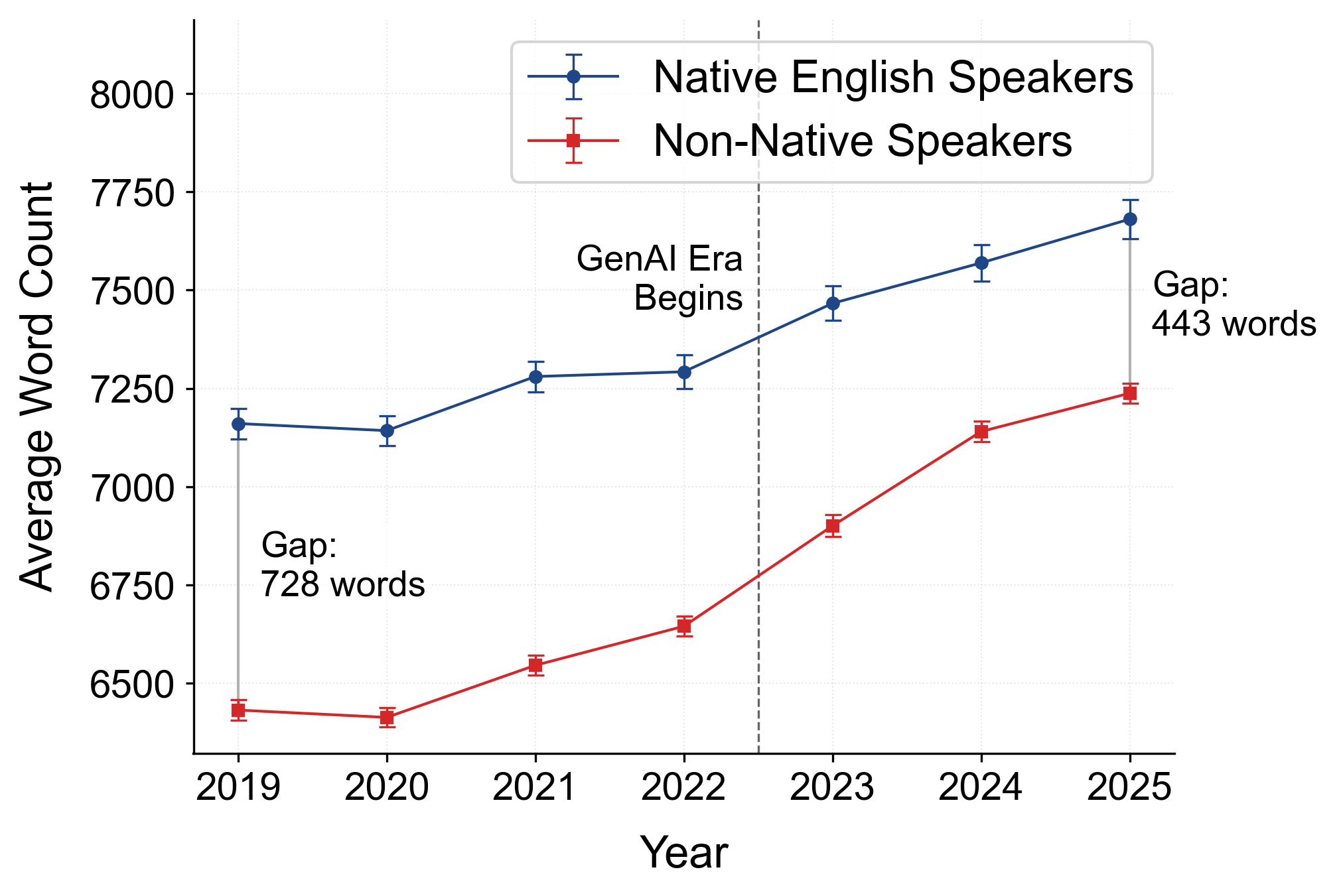}
        \caption{}
        \label{fig:subfig_2a}
    \end{subfigure}
    \begin{subfigure}[b]{0.48\textwidth}
        \centering
        \includegraphics[width=\textwidth]{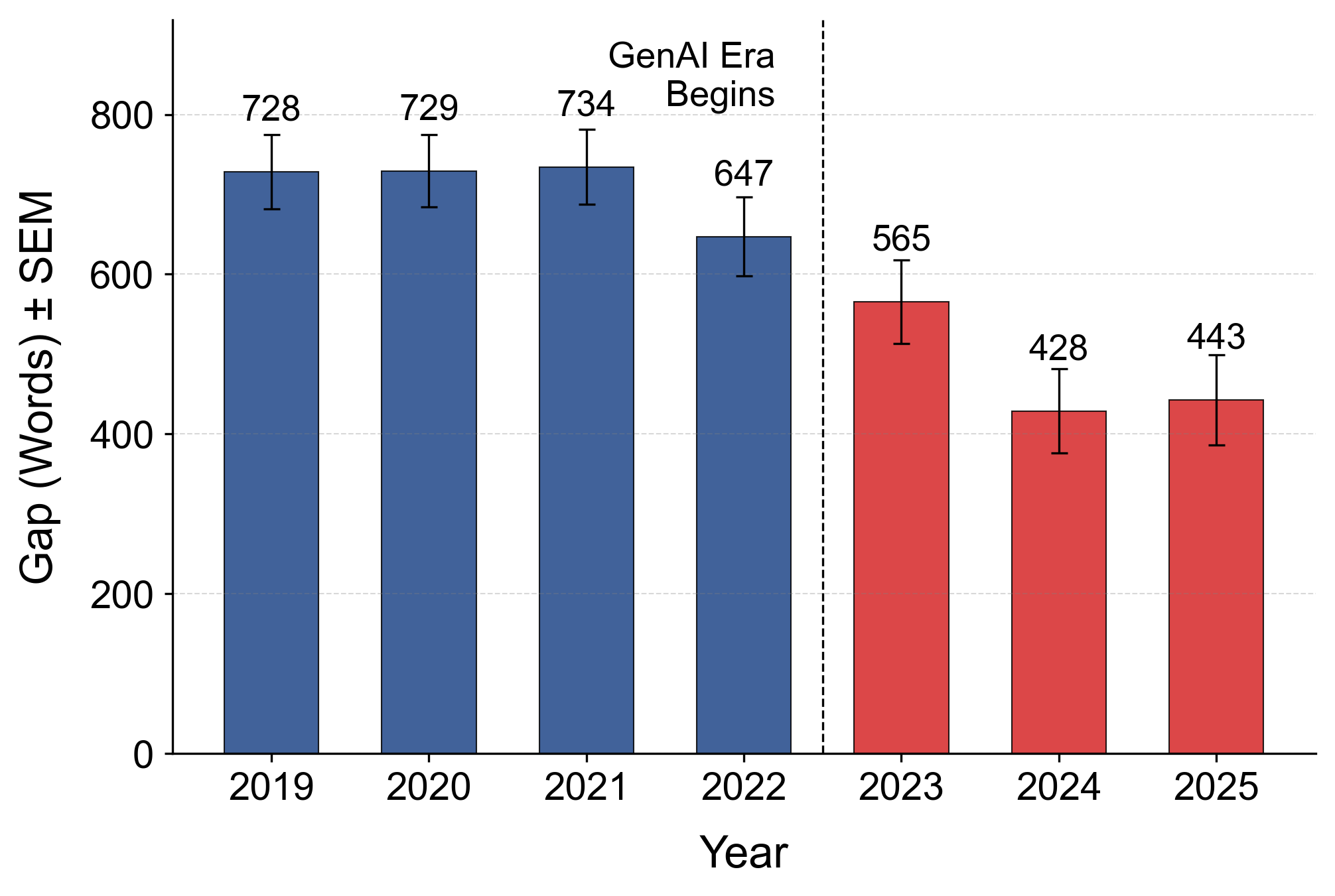}
        \caption{}
        \label{fig:subfig_2b}
    \end{subfigure}
    \caption{\textbf{Convergence of article lengths by author background.} 
    \textbf{(A)} Mean article length (in words, $\pm$ SEM) for NES (blue) and NNES (red) from 2019 to 2025. Following 2022 (vertical dashed line), both groups increased in length, with a larger absolute gain among NNES authors. The gap between groups decreased from 728 words in 2019 to 443 words in 2025.
    \textbf{(B)} Mean difference in word count (± SEM) between NES and NNES across the seven-year period. A stable gap of over 700 words was observed between 2019 and 2022, followed by a significant decrease after 2022. This decline coincides with the onset of the GenAI era (vertical dashed line), with the gap narrowing to its lowest levels in 2024–2025.}
    \label{fig:convergence}
\end{figure}

Figure~\ref{fig:convergence}A shows mean article length by group over time. NES produced longer manuscripts at every time point (all post-hoc $p < .001$). In 2019, the baseline difference was 728 words (7,161 vs.\ 6,432 words). Although the gap remained significant throughout, its magnitude declined overall: the estimated difference on the log$_{10}$ scale decreased from 0.049 in 2019 to 0.028 in 2025.

After 2022, both groups increased in length, but at different rates. NES grew from 7,292 words in 2022 to 7,680 words in 2025 (+5.3\%), whereas NNES increased from 6,645 to 7,238 words (+8.9\%). The raw gap consequently narrowed from 728 words in 2019 to 443 words in 2025, a reduction of 39\%. Figure~\ref{fig:convergence}B presents the mean annual difference in article length between native English-speaking (NES) and non-native English-speaking (NNES) authors.

\subsection{Regional Variations in Article Lengthening (2022 vs.\ 2025)}

To assess regional variation in manuscript lengthening, we compared mean word counts between 2022 and 2025 across major geographic clusters using a two-way ANOVA on log$_{10}$(word count) with year and continent as factors. The analysis revealed significant main effects of year ($F(1, 31{,}931) = 227.17$, $p < .001$) and continent ($F(4, 31{,}931) = 155.53$, $p < .001$), as well as a significant interaction ($F(4, 31{,}931) = 8.90$, $p < .001$), indicating that the magnitude of change differed across regions.

Post-hoc comparisons showed that all five regional groups exhibited significant increases from 2022 to 2025 (all $p < .05$). The largest increase was observed in Africa (Non-Native), rising from 6{,}033 words ($N_{2022} = 1{,}312$) to 6{,}929 words ($N_{2025} = 1{,}227$; +14.8\%; log$_{10}$ difference = 0.058, $p < .001$), followed by Asia (Non-Native), increasing from 6{,}347 ($N_{2022} = 5{,}452$) to 7{,}090 words ($N_{2025} = 8{,}273$; +11.7\%; log$_{10}$ difference = 0.046, $p < .001$). Europe (Non-Native) showed a moderate increase from 7{,}303 ($N_{2022} = 3{,}293$) to 7{,}860 words ($N_{2025} = 2{,}511$; +7.6\%; log$_{10}$ difference = 0.029, $p < .001$). NES increased from 7{,}292 ($N_{2022} = 4{,}796$) to 7{,}680 words ($N_{2025} = 3{,}653$; +5.3\%; log$_{10}$ difference = 0.023, $p < .001$). The smallest increase was observed in America (Non-Native), from 6{,}951 ($N_{2022} = 715$) to 7{,}255 words ($N_{2025} = 709$; +4.4\%; log$_{10}$ difference = 0.022, $p = .018$).

\begin{figure}[htbp]
    \centering
    \includegraphics[width=1.0\textwidth]{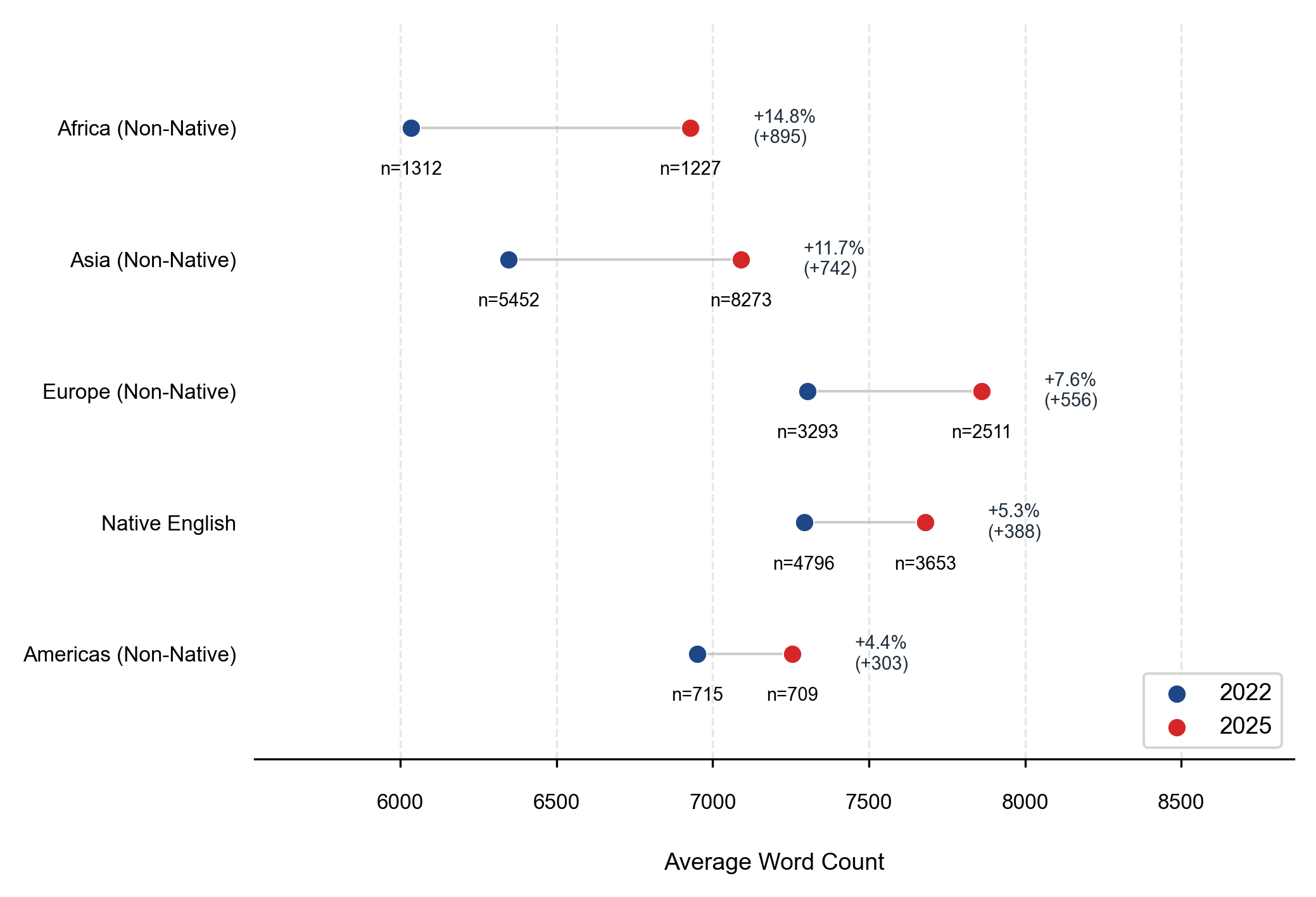}
    \caption{\textbf{Regional differences in manuscript length growth (2022 vs.\ 2025).} 
    Dumbbell plot showing mean word counts in 2022 (blue) and 2025 (red) for each region. 
    Percentages indicate relative growth between the two years, and $n$ denotes the number of manuscripts analyzed per region.}
    \label{fig:regional_growth_22_25}
\end{figure}

Between-group comparisons indicated reduced separation between several regions over time. In 2022, NES differed from Asia (Non-Native) (log$_{10}$ difference = 0.059, $p < .001$), whereas in 2025 the difference was smaller (log$_{10}$ difference = 0.036, $p < .001$). Similarly, Asia (Non-Native) differed from Africa (Non-Native) in 2022 ($p < .001$) but not in 2025 ($p = .413$). Regions with lower baseline word counts exhibited larger post-2022 increases, resulting in reduced cross-regional disparities in average manuscript length by 2025. These regional growth patterns and the resulting reduction in cross-regional disparities are illustrated in Figure~\ref{fig:regional_growth_22_25}.

\subsection{Disciplinary Heterogeneity in Textual Expansion (2022--2025)}

Manuscript lengthening between 2022 and 2025 varied significantly across research fields. A two-way ANOVA on log$_{10}$(word count) with year (2022 vs.\ 2025) and research field as factors revealed significant main effects of year ($F(1, 31{,}893) = 236.50$, $p < .001$) and field ($F(10, 31{,}893) = 188.92$, $p < .001$), as well as a significant interaction ($F(10, 31{,}893) = 10.18$, $p < .001$), indicating that the magnitude of change differed by discipline.

\begin{figure}[h]
    \centering
    \includegraphics[width=\textwidth]{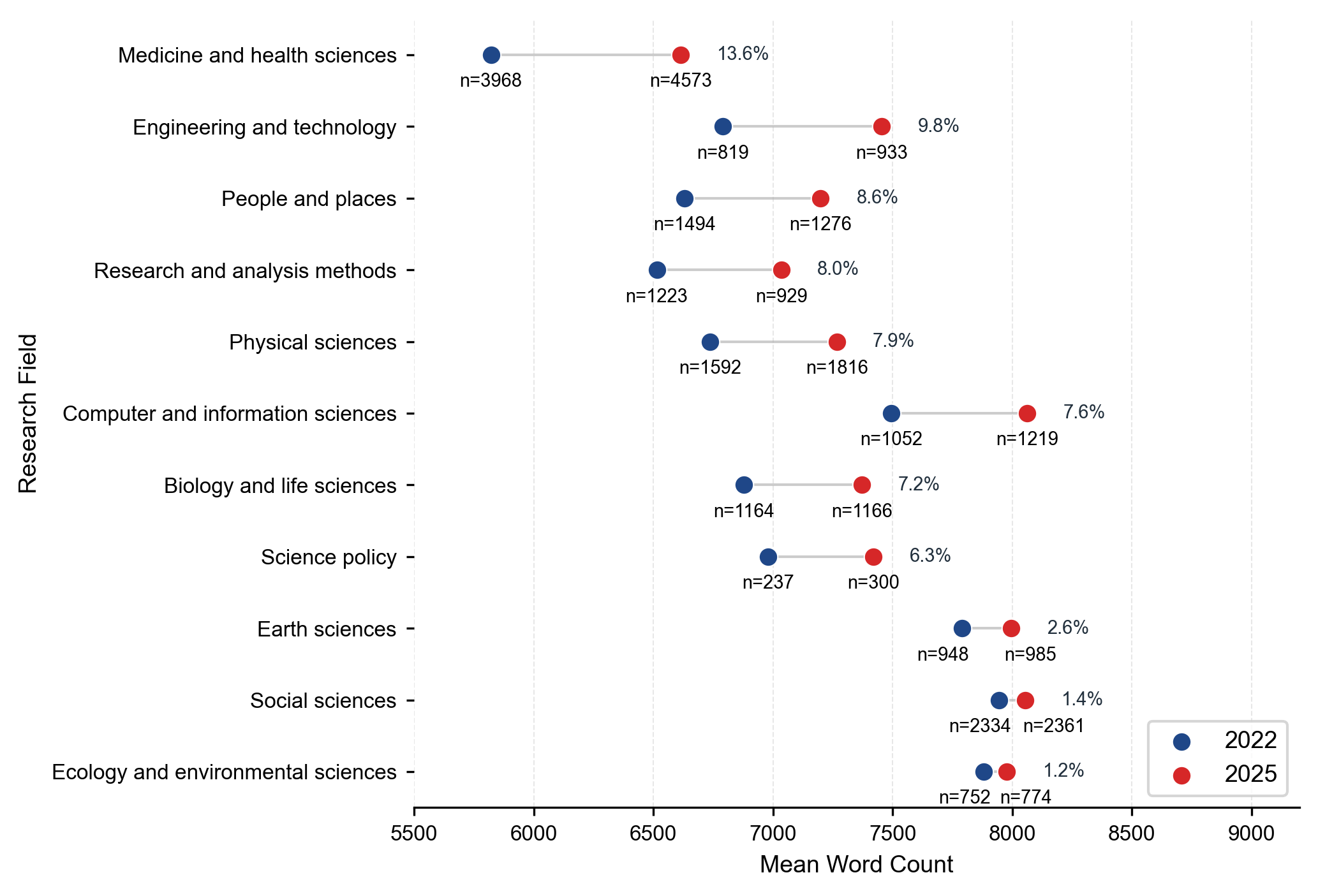}
    \caption{\textbf{Changes in mean article length by research field (2022 vs.\ 2025).} 
    Dumbbell plot showing mean word counts (in words) for 2022 and 2025 across research fields. 
    Points represent annual means, and horizontal segments indicate the difference between years. Percentages displayed to the right of each segment represent the relative growth in word count for that field 
    Sample sizes ($n$) are shown for each year.}
    \label{fig:field_comparison}
\end{figure}

Post-hoc Tukey comparisons showed substantial variation in growth across fields (Figure~\ref{fig:field_comparison}). Medicine and Health Sciences exhibited the largest increase (log$_{10}$ difference = 0.056, $p < .001$; +13.6\%), followed by Engineering and Technology (log$_{10}$ difference = 0.042, $p < .001$; +9.8\%) and People and Places (log$_{10}$ difference = 0.035, $p < .001$; +8.6\%). Significant increases were also observed in Physical Sciences (+7.9\%; $p < .001$), Research and Analysis Methods (+8.0\%; $p < .001$), Biology and Life Sciences (+7.2\%; $p < .001$), and Computer and Information Sciences (+7.6\%; $p < .001$). By 2025, Computer and Information Sciences reached the highest mean word count among all fields, exceeding 8{,}000 words per article.

In contrast, no significant increases were detected in Ecology and Environmental Sciences ($p = .955$), Social Sciences ($p = .776$), Earth Sciences ($p = .465$), or Science Policy ($p = .161$).

\subsection{Difference-in-Differences Regression of Article Length}

We estimated a difference-in-differences (DiD) regression model on log$_{10}$(word count), comparing 2025 to 2022 and incorporating author language background and research field. The model specification was:

\begin{equation}
\log_{10}(\text{Words}_i) = \alpha + \beta_1 \text{Time}_i + \beta_2 \text{Native}_i + \beta_3 (\text{Time}_i \times \text{Native}_i) + \sum \delta_k (\text{Time}_i \times \text{Field}_k) + \varepsilon_i
\end{equation}

\noindent where $\text{Time}$ indicates 2025 (vs.\ 2022), $\text{Native}$ denotes first-author linguistic background, and interaction terms capture differential growth across groups. A total of $N = 31{,}915$ articles were included after excluding records with missing field classifications. Results are reported in Table~\ref{tab:regression}.

Model~1 estimates the unconditional temporal effect between 2022 and 2025 and shows a significant overall increase in article length (log$_{10}$ coefficient = 0.03, $p < .001$; $R^2 = 0.007$).

Model~2 introduces linguistic heterogeneity. NES exhibited higher baseline word counts (0.04 log$_{10}$ units, $p < .001$), corresponding to approximately 9--10\% more words. The interaction term indicates slower growth among NES relative to NNES (interaction = $-$0.01, $p = .005$; $R^2 = 0.014$).

Model~3 incorporates field-specific effects and interactions. The explained variance increases substantially ($R^2 = 0.072$), and the general time effect is reduced to 0.01 and is no longer statistically significant ($p = .161$). Field-specific growth is concentrated in Medicine (0.05, $p < .001$), Engineering (0.04, $p < .001$), and a cluster of fields including Physical Sciences, People-focused Research, Biology, Methods, and Computer Science (all approximately 0.03, all $p < .01$). Policy (0.02, $p = .199$), Earth Sciences (0.00, $p = .695$), and Ecology ($-$0.00, $p = .873$) show no significant field-specific growth.

Importantly, the linguistic interaction remains statistically significant in the full model (interaction = $-$0.01, $p = .040$), indicating that the convergence between NES and NNES authors persists after accounting for disciplinary differences.

\begin{table}[htbp]
\centering
\caption{Hierarchical DiD Regression Results for log$_{10}$(Article Length), 2022--2025}
\label{tab:regression}
\small
\begin{tabular}{lccc}
\hline
\textbf{Variable} & \textbf{Model 1} & \textbf{Model 2} & \textbf{Model 3} \\
\hline
Constant (Baseline) & 3.80*** & 3.79*** & 3.86*** \\
Time (2025 vs.\ 2022) & 0.03*** & 0.04*** & 0.01 \\
NES & -- & 0.04*** & 0.04*** \\
Time $\times$ Native & -- & $-$0.01** & $-$0.01* \\
\hline
\multicolumn{4}{l}{\textit{Baseline Field Differences in 2022 (vs. Social Sciences)}} \\
Medicine & -- & -- & -0.14*** \\
Engineering & -- & -- & -0.07*** \\
Biology & -- & -- & -0.06*** \\
Computer Science & -- & -- & -0.02** \\
Physical Sciences & -- & -- & -0.07*** \\
People-focused Research & -- & -- & -0.08*** \\
Methods & -- & -- & -0.09*** \\
Policy & -- & -- & -0.06*** \\
Earth Sciences & -- & -- & -0.01 \\
Ecology & -- & -- & 0.00 \\
\hline
\multicolumn{4}{l}{\textit{Field-Specific Growth (Interaction with 2025)}} \\
Time $\times$ Medicine & -- & -- & 0.05*** \\
Time $\times$ Engineering & -- & -- & 0.04*** \\
Time $\times$ Biology & -- & -- & 0.03** \\
Time $\times$ Computer Science & -- & -- & 0.03** \\
Time $\times$ Physical Sciences & -- & -- & 0.03*** \\
Time $\times$ People-focused Research & -- & -- & 0.03*** \\
Time $\times$ Methods & -- & -- & 0.03** \\
Time $\times$ Policy & -- & -- & 0.02 \\
Time $\times$ Earth Sciences & -- & -- & 0.00 \\
Time $\times$ Ecology & -- & -- & $-$0.00 \\
\hline
Field Fixed Effects & No & No & Yes \\
Observations & 31,915 & 31,915 & 31,915 \\
$R^2$ & 0.007 & 0.014 & 0.072 \\
\hline
\multicolumn{4}{l}{\footnotesize * $p < .05$, ** $p < .01$, *** $p < .001$. Reference categories: Social Sciences (field), Non-native (language).}
\end{tabular}
\end{table}

\subsection{Authorship Team Size Trends (2019--2025)}

\begin{figure}[ht]
    \centering
    \includegraphics[width=\textwidth]{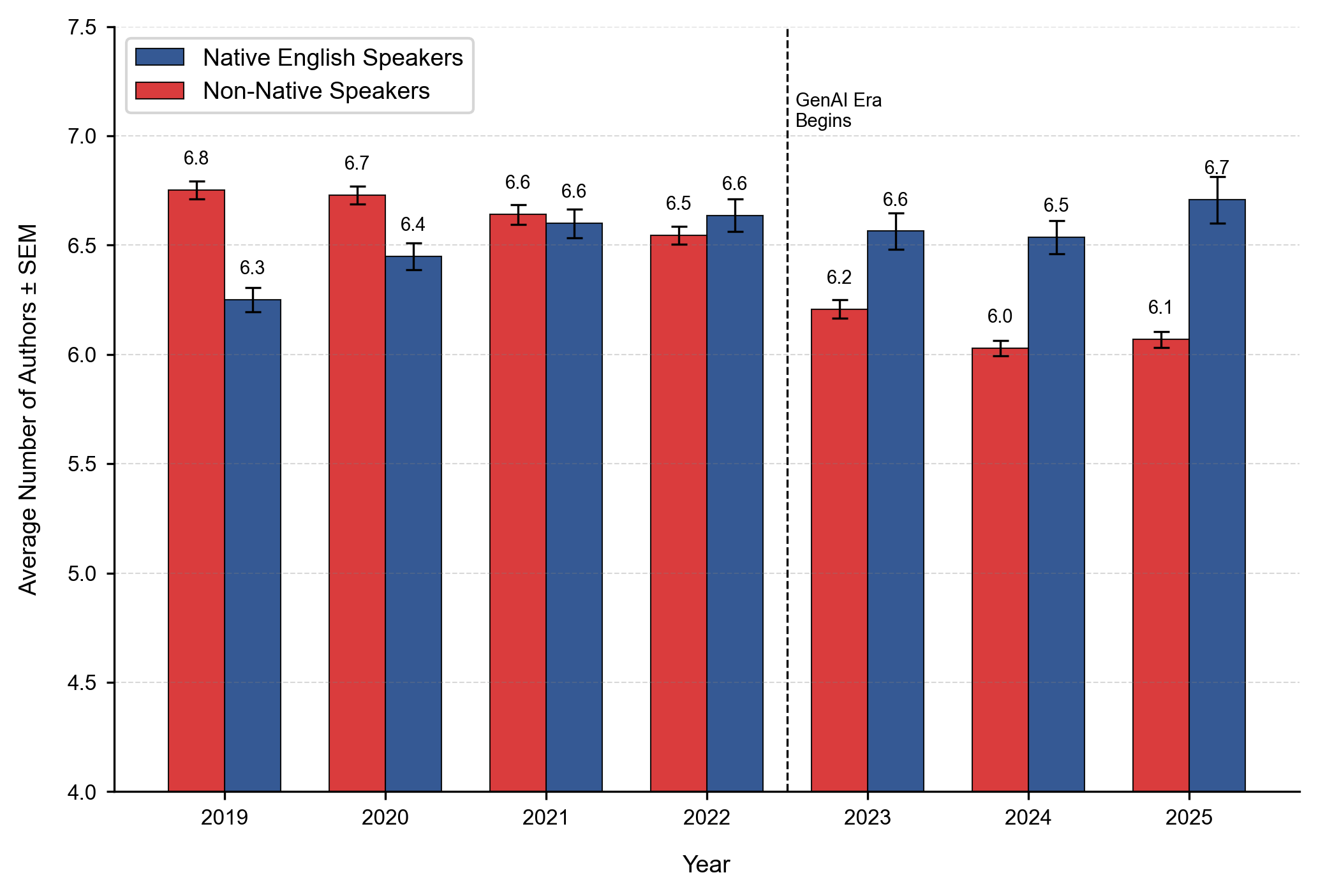}
    \caption{\textbf{Evolution of authorship team size from 2019 to 2025 by first-author linguistic background.} The figure displays the mean number of authors per publication for NES and NNES across the seven-year period. Error bars represent s.e.m. The y-axis is truncated at 4. Statistical comparisons focus on 2022 versus 2025 and reveal a significant interaction ($P < 0.001$), with a reduction in team size observed among NNES authors.}
   \label{fig:authorship}
\end{figure}

A longitudinal analysis of authorship team sizes from 2019 to 2025 revealed a divergence in trends between NES and NNES (Figure~\ref{fig:authorship}). To appropriately model this count outcome, which exhibited overdispersion (variance exceeding the mean), we employed a negative binomial regression with a log link function rather than a standard ANOVA. The model included first-author linguistic background (NES vs.\ NNES), publication year (2022 vs.\ 2025), and their interaction as predictors. Articles with more than 100 authors were excluded, yielding $N = 31{,}963$ observations.

The model revealed a significant interaction between first-author status and year ($\beta = -0.078$, $SE = 0.015$, $z = -5.24$, $p < .001$), indicating that temporal changes in the number of authors differed between NES and NNES first authors. Post-hoc comparisons using estimated marginal means showed that in 2022, there was no significant difference in team size between publications led by NES (mean = 6.61) and NNES (mean = 6.54; ratio = 1.009, $p = .366$). In contrast, by 2025, publications with NES first authors had a significantly higher number of authors (mean = 6.61) compared to those with NNES first authors (mean = 6.06; ratio = 1.091, $p < .001$).

Critically, this divergence was driven by a selective reduction in team size among NNES. For publications with NES first authors, no significant change was observed between 2022 and 2025 (ratio = 0.999, $p = .956$). For NNES first authors, however, the mean number of authors decreased significantly from 6.54 in 2022 to 6.06 in 2025 (ratio = 1.081, $p < .001$). While the longitudinal data in Figure~\ref{fig:authorship} illustrate relative stability in team size for both groups between 2019 and 2022, a distinct separation in trajectories emerged following the 2022 threshold.

\subsection{Post-2022 Shift in International Collaboration}

We analyzed collaboration patterns among 76,338 manuscripts authored by NNES using binary logistic regression, with 2019 as the reference year (Table~\ref{tab:collaboration}). The outcome variable indicated whether at least one English-speaking co-author was present.

Relative to 2019, collaboration declined in 2020 and 2021, partially recovered in 2022, and then decreased substantially from 2023 onward. By 2025, NNES authors were 36\% less likely to collaborate with English-speaking co-authors compared to baseline ($OR = 0.643$, 95\% CI [0.60, 0.69], $p < .001$).

\begin{figure}[htbp]
    \centering
    \begin{subfigure}[b]{0.48\textwidth}
        \centering
        \includegraphics[width=\textwidth]{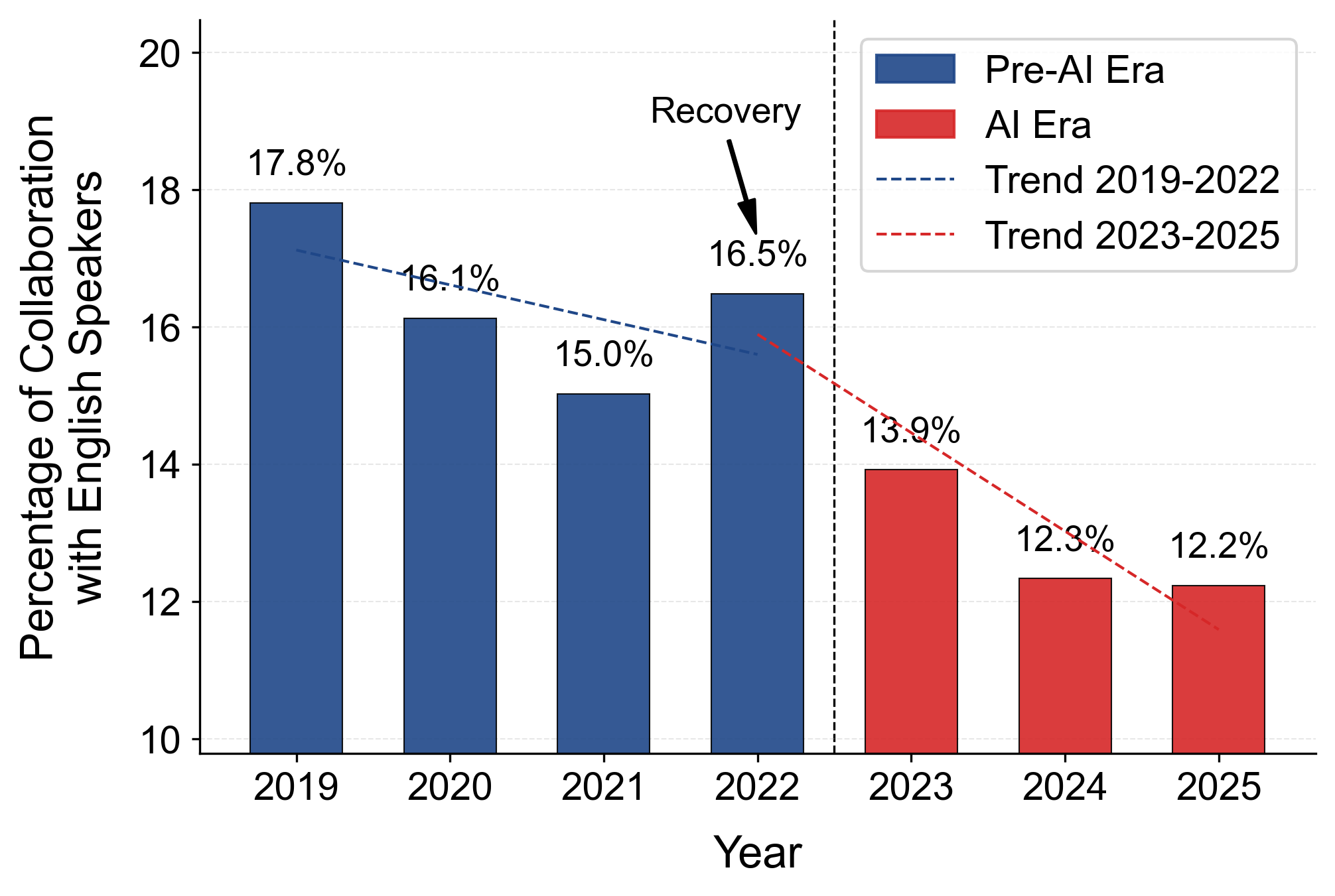}
        \caption{}
        \label{fig:subfig_6a}
    \end{subfigure}
    \begin{subfigure}[b]{0.48\textwidth}
        \centering
        \includegraphics[width=\textwidth]{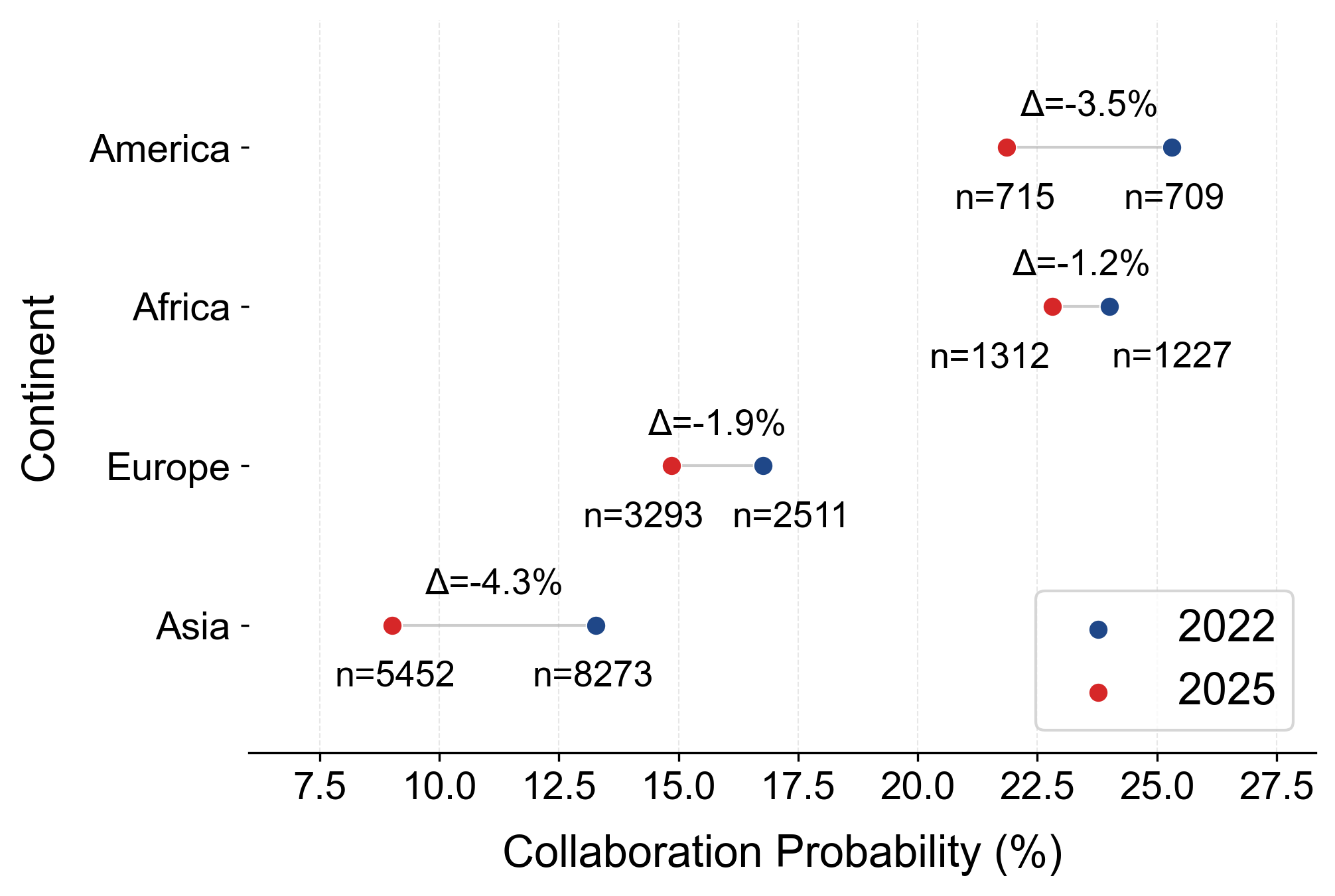}
        \caption{}
        \label{fig:subfig_6b}
    \end{subfigure}
    \caption{\textbf{International collaboration trends (2019--2025).} 
    \textbf{(A)} Year-by-year probability of collaboration with English-speaking co-authors among NNES authors. 
    The figure shows a decline in 2020--2021, partial recovery in 2022, and a marked decrease from 2023 onward.
    \textbf{(B)} The figure displays the probability of collaboration for NNES authors in 2022 and 2025 across different regions. It shows a consistent decrease across all continents by 2025. Percentages above each segment ($\Delta$) indicate the absolute change in probability, and $n$ denotes the sample size per year.}
    \label{fig:collaboration}
\end{figure}

Figure~\ref{fig:collaboration} displays the year-by-year trajectory, highlighting the post-2022 shift in trend.

\begin{table}[htbp]
\centering
\caption{Logistic Regression: Probability of Collaboration with English-Speaking Co-authors}
\label{tab:collaboration}
\small
\begin{tabular}{l c c c}
\hline
\textbf{Year} & \textbf{N} & \textbf{Odds Ratio} & \textbf{95\% CI} \\
\hline
2019 (Baseline) & 10,217 & 1.00  & -- \\
2020 & 10,611 & 0.888 & [0.83, 0.95] \\
2021 & 10,896 & 0.816 & [0.76, 0.88] \\
2022 & 10,788 & 0.911 & [0.85, 0.98] \\
2023 & 9,469  & 0.747 & [0.69, 0.81] \\
2024 & 11,627 & 0.650 & [0.60, 0.70] \\
2025 & 12,730 & 0.643 & [0.60, 0.69] \\
\hline
\multicolumn{4}{l}{\footnotesize Reference year: 2019. Outcome: presence of at least one English-speaking co-author.}
\end{tabular}
\end{table}

\subsection{Reference Count Trends (2022 vs.\ 2025)}

To complement the analysis of textual output, we examined whether the number of references per publication changed differentially between NES and NNES. Given that reference counts are non-negative integers with overdispersion, we employed a negative binomial regression model with a log link function. The dependent variable was the number of references, and the independent variables included first-author linguistic background, publication year (2022 vs.\ 2025), and their interaction ($N = 31{,}364$; 603 observations excluded due to missing values).

The model revealed a significant main effect of first-author status ($\beta = -0.078$, $SE = 0.008$, $z = -9.84$, $p < .001$), with publications by NNES first authors having fewer references on average (incidence rate ratio = 0.93). A significant main effect of year was also observed ($\beta = 0.037$, $SE = 0.010$, $z = 3.72$, $p < .001$), indicating an overall increase in the number of references in 2025 compared to 2022 (incidence rate ratio = 1.04). The interaction between first-author status and year was not statistically significant ($\beta = 0.004$, $SE = 0.012$, $z = 0.31$, $p = .755$), indicating no differential change in reference counts between groups over time.

This null interaction contrasts with the significant differential changes observed for both word count and authorship team size, indicating that the temporal shift documented in textual output did not extend to citation volume.

\section{Discussion}
\subsection{Manuscript Length and Convergence Patterns}

Generative language models can reshape academic writing in two opposing directions. They can condense prose by tightening and refining existing text, or they can reduce the marginal cost of producing additional text, thereby facilitating expansion. Which effect dominates is an empirical question. In our data, the dominant pattern is expansion.

Manuscript length remained statistically stable between 2019 and 2022, with no significant differences in mean or distributional shape. A sharp increase occurred in 2023, representing the largest year over year shift in the study period, followed by additional growth in 2024 and stabilization in 2025. Although this timing aligns with the diffusion of generative AI writing tools, temporal coincidence alone is insufficient for causal inference. The design is observational. The inclusion of field fixed effects and year by field interactions reduces concerns about compositional shifts across disciplines, but concurrent structural influences cannot be fully excluded.

The aggregate increase conceals meaningful heterogeneity. Between 2022 and 2025, NES authors increased manuscript length by 5.3\%, whereas NNES authors increased by 8.9\%. The word count gap, which had remained approximately 730 words between 2019 and 2021, narrowed to 442 words by 2025, representing a 39\% reduction. The strongest compression occurred in 2023 and 2024, followed by stabilization.

The convergence follows a clear gradient. Groups with lower baseline word counts exhibit larger subsequent increases. African affiliated first authors show the largest growth of 14.8\%, followed by Asian affiliated authors at 11.7\%, European NNES authors at 7.6\%, and NES authors at 5.3\%. A comparable ordering appears across disciplines, with Medicine and Health Sciences and Engineering showing larger increases than Social Sciences and Ecology.

\citet{lin2025chatgpt} document substantial post 2022 variation in lexical shifts across fields, with the strongest effects observed in technology related disciplines and biology, and no significant effects in Social Sciences or Arts and Humanities. This disciplinary heterogeneity suggests that adoption or effective integration of generative tools may itself be field dependent. \citet{ding2025rise} similarly document that the expansion of generative AI research after 2022 has been highly uneven across disciplines, with particularly rapid growth in computer science and related technological domains. If so, part of the variation observed here may reflect differences in the speed or intensity of uptake across disciplinary communities, rather than purely linguistic constraints.

At the same time, regression to the mean could mechanically generate stronger growth among groups starting from lower baselines. The persistence of the interaction term in the difference in differences specification, even after controlling for field fixed effects and year by field interactions, indicates that convergence is not solely a statistical artifact. In the full model, the main time effect becomes statistically negligible, while the interaction between year and linguistic background remains statistically significant (p = .040). Manuscript lengthening therefore reflects both field specific dynamics and linguistic background effects rather than simple compositional redistribution across disciplines. Nonetheless, the design does not establish causality.

These structural patterns align with recent evidence of post 2022 convergence in other textual dimensions. \citet{lin2025chatgpt} report increases in lexical complexity among NNES authored abstracts in large scale OpenAlex data, while \citet{filimonovic2025generative} document stylistic convergence toward NES writing conventions across millions of articles. Similarly, \citet{bao2025examining} identify post-ChatGPT shifts in lexical complexity, syntactic simplification, and stronger stylistic change among NNES authors in arXiv abstracts. Although these studies examine vocabulary and stylistic features rather than length, the parallel movement across lexical, stylistic, and structural dimensions suggests that the post 2022 period may be associated with multi dimensional shifts in academic writing, particularly among authors facing higher initial linguistic barriers.

Recent evidence also indicates geographic heterogeneity in AI related academic behavior. \citet{ben2026search} document sharper declines in traditional academic search activity in non English speaking regions compared to the United States and the United Kingdom following the introduction of generative tools. Although their study examines information consumption rather than manuscript production, the geographic gradient they identify is consistent with the possibility that differential uptake contributes to the patterns observed here. This interpretation remains suggestive rather than causal.

Increased manuscript length may reflect more thorough reporting, expanded methodological transparency, or richer argumentation. It may also reflect a reduction in the marginal cost of textual production without proportional gains in substantive contribution. Two distinct questions follow. First, what mechanisms drive differential adoption across linguistic and disciplinary groups? Second, does textual expansion translate into measurable improvements in scientific quality or impact? The former concerns incentive structures and constraints in scholarly production. The latter is empirically testable through linkage to downstream indicators such as citation impact and related measures of scholarly influence.

\subsection{Authorship Team Size and International Collaboration}

Beyond textual expansion, the diffusion of generative language models may influence the social organization of scientific production. If such tools reduce the cost of drafting, revising, and refining manuscripts in English\citep{aydin2023main}, they may partially substitute for contributions previously provided by collaborators who served linguistic functions.

The results indicate a shift in team composition after 2022. In 2022, there was no statistically significant difference in mean team size between publications led by NES and NNES first authors (6.61 versus 6.54 authors, $p = .366$). By 2025, a statistically significant difference had emerged: publications led by NES first authors remained stable in size (mean = 6.61), whereas those led by NNES first authors declined to a mean of 6.06 authors ($p < .001$). The reduction between 2022 and 2025 is observed exclusively among NNES first authors.

A parallel pattern appears in international collaboration. Restricting the analysis to publications with NNES first authors, we modeled the probability of including at least one NES co-author. Relative to 2019 as the reference year, collaboration declined in 2020 and 2021, partially recovered in 2022, and then decreased sharply and consistently from 2023 onward. By 2025, NNES first authors were 36\% less likely to collaborate with a NES co-author compared to 2019 (OR = 0.643, $p < .001$).

The 2020 and 2021 decline coincides with COVID-19-related disruptions to international collaboration \citep{cai2021international, abramo2022covid,ni2025navigating}, and the 2022 recovery aligns with resumed academic mobility. The renewed decline beginning in 2023 occurs in a different temporal context and is unlikely to be fully explained by pandemic aftereffects alone. Alternative explanations warrant consideration. 

Policy-driven geopolitical tensions may plausibly contribute to the observed decline in cross-linguistic collaboration. Bibliometric analyses of the U.S. "China Initiative," launched in 2018 and formally terminated in 2022, indicate a substantial reduction in U.S.–China co-authorship, accompanied by measurable spillover effects across Western collaboration networks \citep{geng2025displacement}. These reported estimates are consistent with a displacement dynamic, in which reduced collaboration with U.S. and G6 partners coincided with increased domestic collaboration within China. Such a reallocation of collaborative ties would be expected to modify the composition of international networks rather than induce a uniform reduction in collaboration rates across all geographic regions. Two further patterns complicate a purely geopolitical account. First, the temporal trajectory is inconsistent with policy-driven suppression: 2019, the year following the China Initiative’s launch, recorded the highest rate of cross-linguistic collaboration in the entire study period (17.8\%). Second, and most critically, the post-2022 contraction is not confined to the Sino-American axis; as shown in our regional analysis (Figure~\ref{fig:collaboration}), the decline is observed consistently across Asia, Europe, Africa, and Latin America. This universal trend suggests a global rather than a localized mechanism. While direct causality cannot be established from bibliometric trends alone, the concurrent contraction in team size, the decline in native-speaker collaboration, and the differential expansion in manuscript length among NNES authors are consistent with the potential impact of generative writing tools. These findings point toward a plausible technology-driven reorganization of scientific co-authorship, a hypothesis that warrants further investigation through direct surveys or stylistic text analysis.

Prior research links cross-national collaboration to cognitive diversity, broader epistemic perspectives, and higher scientific impact \citep{wang2024elevating,velez2022papers,lariviere2015team}. If collaboration with NES co-authors previously fulfilled linguistic drafting or editing functions alongside substantive scientific contribution \citep{lillis2006professional}, generative tools may now perform part of this role. The selective reduction in team size among NNES first authors, combined with the concurrent decline in NES collaboration, is consistent with a possible reorganization of the functional role that linguistic collaboration previously played. This interpretation remains speculative. The analysis does not establish causality, and tool use is not directly observed. Continued longitudinal analysis will be necessary to determine whether these patterns reflect a temporary adjustment or a durable structural shift.

\subsection{Reference Count Trends}
Reference counts present a contrasting pattern. Between 2022 and 2025, the number of references per article increased modestly (incidence rate ratio = 1.04, $p < .001$), but this increase was uniform across NES and NNES authors, with no significant interaction ($p = .755$). This null interaction serves as an internal benchmark: whereas manuscript length and collaboration patterns show differential shifts by linguistic background, bibliographic output does not.

The contrast suggests that the post-2022 changes documented above are specific to dimensions of writing where generative tools most directly reduce production costs, namely text generation and language refinement, rather than reflecting a general inflation across publication indicators. Bibliographic integration, which requires identifying relevant sources and situating findings within an existing literature, may be less directly affected by generative tools than prose composition.

At the same time, concerns about fabricated references in LLM-generated text \citep{chelli2024hallucination} make the absence of differential change noteworthy: it suggests that any adoption of generative tools among NNES authors has not produced detectable distortions in citation practices at the aggregate level.

\section{Limitations}

Several limitations of the present study warrant consideration.

First, the analysis is restricted to a single journal, PLOS ONE. Although this design 
offers the methodological advantage of a consistent editorial framework across 
disciplines and time, it limits the generalizability of the findings. Replication 
across journals of varying selectivity is needed before broader conclusions can be drawn.

Second, the study identifies temporal associations between post-2022 trends and 
structural publication characteristics but cannot establish causal attribution. No 
direct measure of LLM usage is available at the document level, and no external 
control condition exists.

Third, the classification of authors as NES or NNES relies 
on institutional affiliation rather than on the author's actual linguistic background. 
This operationalization introduces systematic misclassification: researchers whose 
first language is not English but who are affiliated with institutions in 
English-majority countries are coded as NES speakers, and vice versa.

This misclassification is, however, likely to attenuate rather than 
inflate the observed effects: NNES researchers embedded in 
English-majority institutions are coded as NES, reducing measured 
differences between groups. The reported estimates should therefore 
be interpreted as conservative lower bounds on the true magnitude 
of the documented patterns.

Fourth, the observed decline in collaboration with NES co-authors may reflect, in part, policy-level constraints on international research collaboration in certain countries, most notably China, which represents a substantial share of Asian-affiliated authors in our sample. We are unable to quantify the extent of this contribution, and disentangling policy-driven effects from technology-driven ones remains a direction for future research

\section{Conclusion}

This study documents three concurrent post-2022 shifts in scientific publishing: manuscripts grew longer, the word-count gap between NES and NNES authors narrowed, and NNES authors reduced both their team sizes and their collaboration with NES co-authors. The first two findings complement existing evidence on lexical and stylistic convergence; the third has not been previously documented. This finding is consistent with the interpretation that generative language tools may be reshaping not only how scientific texts are produced, but how collaborative structures are organized, partially substituting for relationships that previously served a linguistic function. Across all measures, the largest effects were observed among the populations where pre-existing gaps were widest, including authors affiliated with institutions in Africa and Asia and those in disciplines with lower baseline manuscript lengths, consistent with a technology meeting an existing need.

\citep{lepori2025generative} argue that if generative AI changes how scientific texts are written, then common scientometric indicators such as manuscript length, authorship patterns, and references may no longer mean exactly what they used to. The changes documented here are consistent with this concern, and it is possible that we are already in the midst of a shift in how scientific texts are produced, one that may carry implications for how standard publication metrics are interpreted. If so, part of the post 2022 movement in publication measures may reflect changes in the production of scientific text rather than changes in the underlying substance of scientific research.

Whether this reorganization represents a net benefit for the global scientific system remains an open question. The narrowing of language-related disparities points toward broader access; the decline in cross-linguistic collaboration points toward greater insularity. These trajectories are not necessarily in tension. Reduced dependence on linguistic mediation may, over time, enable new collaborations grounded in scientific rather than language-based complementarity. However, the present data capture only the initial phase of this transition, and its longer-term implications require continued longitudinal examination.

\subsection*{Author Contributions}
Conceptualization: Y. Ben-Zion; Methodology: Y. Ben-Zion; 
Formal analysis and investigation: Y. Ben-Zion, E. Cohen; 
Writing - original draft preparation: Y. Ben-Zion, E. Cohen, N. Davidovitch; 
Writing - review and editing: Y. Ben-Zion, E. Cohen, N. Davidovitch; 
Supervision: Y. Ben-Zion.

\subsection*{Funding}
The authors declare that no funds, grants, or other support were received during the preparation of this manuscript.

\subsection*{Competing Interests}
The authors have no relevant financial or non-financial interests to disclose.

\subsection*{Availability of data and materials}
The dataset supporting the findings of this study is openly available on Zenodo at https://doi.org/10.5281/zenodo.18790135

\bibliography{sn-bibliography}

\section*{Appendix A: Classification Prompts}

Both prompts were designed through an iterative process of trial and error 
to accommodate the specific structure of PLOS ONE XML records, and in 
particular the bracket-based affiliation indexing system unique to this 
journal. Each prompt was tested and refined until it produced accurate and 
consistent classifications. All model outputs were subject to manual 
inspection, and the final prompt formulations reported here are those that 
yielded the classifications used in the analysis. Articles were submitted 
to the model in batches, a standard approach for classification tasks of 
this scale.

\subsection{A.1 Native English Speaker Classification}

The following prompt was submitted to the GPT-5 mini API.

\begin{quote}
\texttt{You will receive multiple articles (up to 150). For EACH article, 
analyze the author affiliations independently.}\\

\texttt{For each article's affiliation list:}\\
\texttt{- Each affiliation list is separated from the other by |.}\\
\texttt{- Each list has authors separated by commas (,).}\\
\texttt{- The institutional affiliation code appears in square brackets [] 
after each author's name.}\\
\texttt{- The same code appears before the institutional affiliation itself.}\\

\texttt{IMPORTANT: The institutional affiliation codes are unique PER 
ARTICLE. Different articles can have the same code (e.g., [aff001]) but 
refer to different institutions!}\\

\texttt{For each article, answer these 3 questions:}\\
\texttt{Question 1: Is the FIRST author from a native-English-majority 
country?}\\
\texttt{Question 2: Is the LAST author from such a country?}\\
\texttt{Question 3: Is at least ONE author from such a country?}\\

\texttt{Native-English-majority countries: USA, UK, Ireland, Australia, 
New Zealand, Canada.}\\

\texttt{IMPORTANT: If an author has multiple affiliations (e.g., 
[aff001, aff002]), check ALL of them. If at least ONE is from a qualifying 
country, the answer is YES (1).}\\

\texttt{Example:}\\
\texttt{Article 1:}\\
\texttt{Authors: Michelle Da Silva Lodge [aff001] | Nick Pullen [aff002] 
| Miguel Pereira [aff003] | Timothy S. Johnson [aff002]}\\
\texttt{Affiliations: [aff001] Academic Nephrology Unit, University of 
Sheffield, UK | [aff002] Pfizer Global R\&D, Cambridge, MA, USA | 
[aff003] UCB Pharma, Slough, UK}\\
\texttt{First author: Michelle [aff001] → UK → 1}\\
\texttt{Last author: Timothy [aff002] → USA → 1}\\
\texttt{At least one: Yes → 1}\\
\texttt{Output: 1 1 1}\\

\texttt{OUTPUT FORMAT:}\\
\texttt{Return ONLY a JSON object with this structure:}\\
\texttt{"results":}\\
\texttt{ \{"article": 1, "analysis": "1 1 1"\}}\\
\texttt{Do NOT include any other text, explanations, or markdown 
formatting. ONLY the JSON object.}\\

\texttt{Here are the articles to analyze:}
\end{quote}

\subsection{A.2 Geographic (Continent) Classification}

The following prompt was submitted to the GPT-5 mini API.

\begin{quote}
\texttt{You will receive multiple articles (up to 150). For EACH article, 
analyze the author affiliations independently.}\\

\texttt{For each article's affiliation list:}\\
\texttt{- Each affiliation list is separated from the other by |.}\\
\texttt{- Each list has authors separated by commas (,).}\\
\texttt{- The institutional affiliation code appears in square brackets [] 
after each author's name.}\\
\texttt{- The same code appears before the institutional affiliation itself.}\\

\texttt{IMPORTANT: The institutional affiliation codes are unique PER 
ARTICLE. Different articles can have the same code (e.g., [aff001]) but 
refer to different institutions!}\\

\texttt{For each article, answer these questions:}\\
\texttt{Question 1: Which continent is the FIRST author from?}\\
\texttt{Question 2: Which continent is the LAST author from?}\\
\texttt{Question 3: For EACH continent, is there AT LEAST ONE author from 
that continent? Answer with 5 binary digits (0 or 1) in this exact order:}\\
\texttt{Position 1: Is there at least one author from Asia? (1=yes, 0=no)}\\
\texttt{Position 2: Is there at least one author from Europe? (1=yes, 0=no)}\\
\texttt{Position 3: Is there at least one author from Africa? (1=yes, 0=no)}\\
\texttt{Position 4: Is there at least one author from America? (1=yes, 0=no)}\\
\texttt{Position 5: Is there at least one author from Other regions? 
(1=yes, 0=no)}\\

\texttt{Continent codes:}\\
\texttt{1 = Asia (China, Japan, India, Israel, South Korea, Singapore, 
Thailand, Malaysia, Indonesia, etc.)}\\
\texttt{2 = Europe (UK, Germany, France, Spain, Italy, Netherlands, 
Switzerland, Sweden, Norway, Poland, etc.)}\\
\texttt{3 = Africa (South Africa, Egypt, Nigeria, Kenya, Ethiopia, 
Morocco, etc.)}\\
\texttt{4 = America (USA, Canada, Brazil, Mexico, Argentina, Chile, 
Colombia, etc. - all of North and South America)}\\
\texttt{5 = Other (Australia, New Zealand, and any other regions)}\\
\texttt{0 = Unknown or cannot determine}\\

\texttt{IMPORTANT: If an author has multiple affiliations (e.g., 
[aff001, aff002]), check ALL of them. If affiliations are from different 
continents, use the FIRST continent mentioned.}\\

\texttt{Example:}\\
\texttt{Article 1:}\\
\texttt{Authors: Michelle Da Silva Lodge [aff001] | Nick Pullen [aff002] 
| Miguel Pereira [aff003] | Timothy S. Johnson [aff002]}\\
\texttt{Affiliations: [aff001] Academic Nephrology Unit, University of 
Sheffield, UK | [aff002] Pfizer Global R\&D, Cambridge, MA, USA | 
[aff003] UCB Pharma, Slough, UK}\\
\texttt{First author: Michelle [aff001] → UK → Europe → 2}\\
\texttt{Last author: Timothy [aff002] → USA → America → 4}\\
\texttt{Presence by continent: Asia=0, Europe=1 (Michelle, Miguel), 
Africa=0, America=1 (Nick, Timothy), Other=0 → 01010}\\
\texttt{Output: 2 4 01010}\\

\texttt{Article 2:}\\
\texttt{Authors: Li Wang [aff001] | Sarah Cohen [aff002]}\\
\texttt{Affiliations: [aff001] Peking University, Beijing, China | 
[aff002] Tel Aviv University, Israel}\\
\texttt{First author: Li [aff001] → China → Asia → 1}\\
\texttt{Last author: Sarah [aff002] → Israel → Asia → 1}\\
\texttt{Presence by continent: Asia=1 (both authors), Europe=0, Africa=0, 
America=0, Other=0 → 10000}\\
\texttt{Output: 1 1 10000}\\

\texttt{OUTPUT FORMAT:}\\
\texttt{Return ONLY a JSON object with this structure:}\\
\texttt{results:}\\
\texttt{\{"article": 1, "analysis": "2 4 01010"\}}\\
\texttt{\{"article": 2, "analysis": "1 1 10000"\}}\\
\texttt{IMPORTANT: Question 3 must always contain exactly 5 binary digits 
(0 or 1) representing [Asia][Europe][Africa][America][Other] in that order.}\\
\texttt{Do NOT include any other text or explanations.}\\

\texttt{Here are the articles to analyze:}
\end{quote}

\end{document}